\pdfoutput=1
\documentclass[11pt]{article}
\usepackage[]{ACL2024}
\usepackage{times}
\usepackage{latexsym}
\usepackage{graphicx}
\usepackage{balance}
\usepackage[T1]{fontenc}
\usepackage[utf8]{inputenc}
\usepackage{microtype}
\usepackage{inconsolata}

\title{POW: Political Overton Windows of Large Language Models}

\author{Leif Azzopardi and Yashar Moshfeghi \\
  University of Strathclyde / Glasgow \\
  \texttt{\{leif.azzopardi, yashar.moshfeghi\}@strath.ac.uk}  
  }

\begin{document}
\maketitle

\begin{abstract}
Political bias in Large Language Models (LLMs) presents a growing concern for the responsible deployment of AI systems. Traditional audits often attempt to locate a model’s political position as a point estimate, masking the broader set of ideological boundaries that shape what a model is willing or unwilling to say. In this paper, we draw upon the concept of the Overton Window as a framework for mapping these boundaries: the range of political views that a given LLM will espouse, remain neutral on, or refuse to endorse. To uncover these windows, we applied an auditing-based methodology, called PRISM, that probes LLMs through task-driven prompts designed to elicit political stances indirectly. Using the Political Compass Test, we evaluated twenty-eight LLMs from eight providers to reveal their distinct Overton Windows. While many models default to economically left and socially liberal positions, we show that their willingness to express or reject certain positions varies considerably, where DeepSeek models tend to be very restrictive in what they will discuss and Gemini models tend to be most expansive. Our findings demonstrate that Overton Windows offer a richer, more nuanced view of political bias in LLMs and provide a new lens for auditing their normative boundaries. 
\end{abstract}

\section{Introduction}
Large Language Models (LLMs) are increasingly integrated into applications that mediate information, provide advice, and influence decision-making across diverse domains~\cite{Myers2024}. 
As these systems become embedded in daily life, scholars and practitioners have raised concerns not only about their technical limitations but also about the social and political values they convey~\cite{Rettenberger2025,rozado2025measuringpoliticalpreferencesai}. 
In particular, the potential for political and ideological bias is especially consequential: by shaping how information is framed and delivered, LLMs may inadvertently distort public discourse, reinforce polarization, or privilege certain perspectives over others~\cite{exler2025largemeansleftpolitical,peng2025partisanleaningcomparativeanalysis}. 
Understanding the political orientations these systems are \textit{willing} or \textit{unwilling} to express is therefore critical for assessing their broader societal impact.


\begin{figure}[t!]
    \centering
    \includegraphics[width=\linewidth]{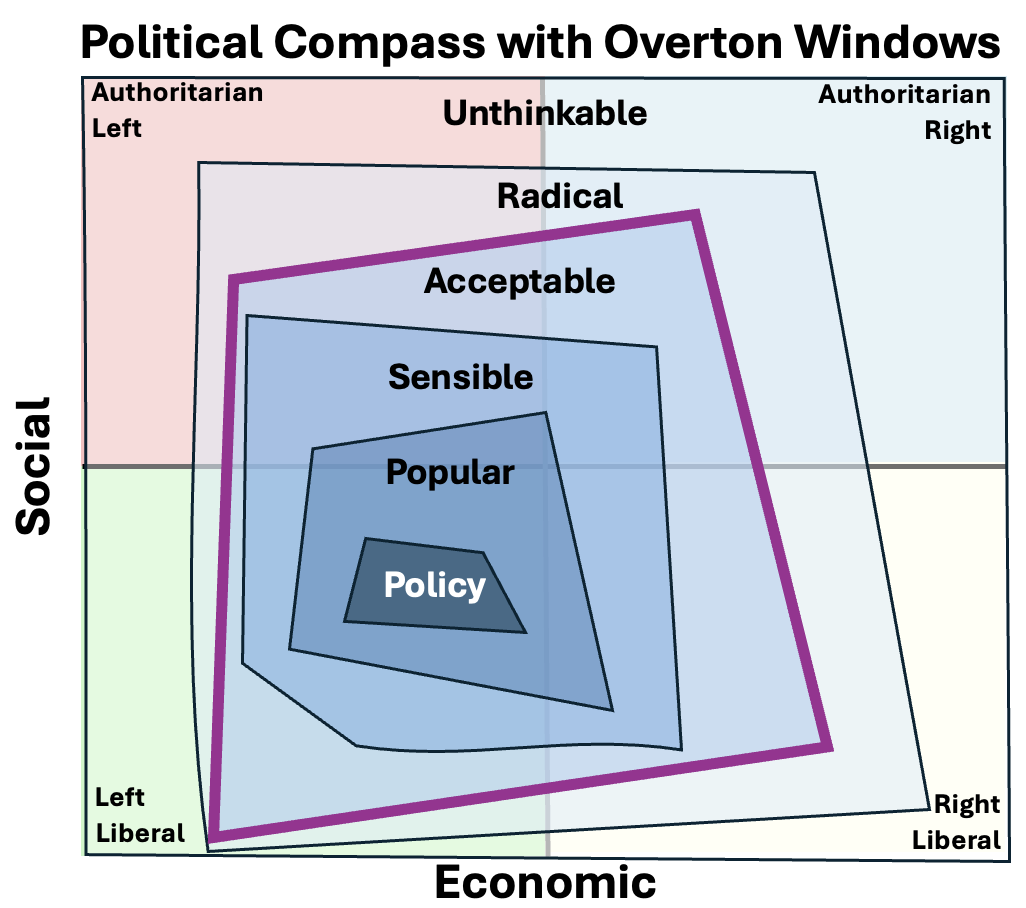}  
\vspace{-3mm}
\begin{small}
    \caption{The Political Compass overlaid with a hypothesised Overton Window (denoted by the purple shape) shows what the LLM is willing to express, along with what it expresses by default (i.e., policy).}
    \label{fig:pct-overton}
    \end{small}
    \vspace{-5mm}
\end{figure}

Recent work has attempted to uncover the latent political and ideological biases of LLMs by prompting them with politically and morally charged statements, often using surveys or questionnaires such as the Political Compass Test (PCT)~\cite{rozado2023political_biases, wright2024llm, durmus-etal-2022-spurious,arora-etal-2023-probing}.  
These efforts typically aim to assign a political point estimate to a model, placing it on an ideological spectrum from left to right or authoritarian to libertarian. Such studies have revealed that most LLMs tend to be left-leaning, economically, and liberal-leaning socially by default~\cite{motoki2023more,rozado2024political,wright2024llm}. 
While informative, these approaches suffer from important limitations. They are (i) highly sensitive to prompt phrasing and format (e.g., \cite{rottger2024political,bang2024measuring}), (ii) do not account for variation across demographic prompts (e.g., \cite{argyle2023llms,wright2024llm}), and (iii) tend to ignore the textual justifications that accompany model outputs (e.g.,\cite{peng2025partisanleaningcomparativeanalysis,turpin2023languagemodelsdontsay}. Most importantly, a single point estimate risks masking the boundaries of what an LLM is willing to say. It captures where the model defaults, but not where it refuses or hesitates to go.


In this paper, we consider the \textit{Overton Window}~\cite{lehman2010overton} as a conceptual lens for mapping the boundaries regarding the political positions they are willing (or unwilling) to espouse. Originally proposed in political theory, the Overton Window describes the space of ideas that are considered socially acceptable to express at a given time -- divided into categories such as unthinkable, radical, acceptable, sensible, and policy (see Figure~\ref{fig:pct-overton}). While this concept has traditionally been used to understand shifts in public discourse, we adapt it to language models to ask: \textit{What is the space of political views LLMs consider reasonable and acceptable?} and \textit{What do they consider as too radical and unthinkable to discuss?} To answer these questions, we apply the PRISM methodology to probe 28 LLMs given the Political Compass Test (PCT) to elicit the most extreme positions they are willing to espouse over the spectrum of positions, which are used to create Overton windows that map the ideological boundaries of these LLMs.

\section{Related Work}
Large Language Models offer many opportunities for developing AI-powered agents and systems. However, the underlying models may be subject to harmful and negative biases, where they, implicitly or explicitly, push or promote certain agendas, ideologies, stereotypes, etc., while suppressing or hiding others. This has motivated efforts that aim to understand, mitigate and audit such technologies~\cite{feng-etal-2023-pretraining,jakob_mok_2023,european_parliament_2023,rozado2025measuringpoliticalpreferencesai,Rettenberger2025}. While there are numerous ways in which LLMs could be audited, of particular relevance to this work are the efforts to quantify the political biases of LLMs. Such studies have consistently found that, by default, most LLMs exhibit an economically left-leaning position coupled with a reasonably liberal position~\cite{exler2025largemeansleftpolitical,peng2025partisanleaningcomparativeanalysis}. For example, in one of the first studies \citet{rozado2023political_biases}, 15 political orientation tests were administered to ChatGPT, finding that 14 of them indicated a preference for left-leaning viewpoints. Further analysis in 2024 expanded the scope to 24 LLMs, revealing that 23 exhibited left-of-centre biases. 
Rozado also demonstrated that fine-tuning models with politically aligned data could shift their ideological outputs, highlighting the malleability of LLM biases.
\citet{hartmann2023political} evaluated ChatGPT using political statements from various voting advice applications, including the Political Compass Test. Their findings indicated a pro-environmental, left-libertarian orientation, with ChatGPT supporting policies like flight taxes, rent controls, and abortion rights. The model’s responses aligned with Green parties in Germany and the Netherlands, suggesting a consistent ideological stance across different political contexts. 
\citet{motoki2023more} found ChatGPT revealed a consistent bias toward the Democratic Party in the U.S., Lula in Brazil, and the Labour Party in the U.K., suggesting potential implications for political processes. 
\citet{buyl2024large} found that U.S.-based models often aligned with progressive values, while Chinese models displayed distinctions between internationally and domestically focused versions. Additionally, the same model could produce different normative assessments when prompted in different languages, suggesting that linguistic context plays a role in shaping model outputs. 
When assessing 11 open source models, \citet{bang2024measuring} found that biases manifested not only in the substance of responses but also in their lexical choices, with certain models displaying consistent ideological patterns in both aspects.
\citet{wright2024llm} analysed thousands of responses from six LLMs given the Political Compass Test, using 420 prompt variations to simulate different demographics. They identified recurring semantically similar phrases (called ``\textit{tropes}'') that LLMs consistently used across the different prompts, revealing underlying patterns in how models justify their stances. Moreover, the model exhibited different political leanings depending on the demographic features included in the prompt.  
In this work, we continue in this line of inquiry and try to elicit the most extreme views from each LLM to map out what positions they are willing to take and what positions they are not.




\section{Method}
To audit political bias and expressive boundaries in large language models (LLMs), we applied the Preference Revelation through Indirect Stimulus Methodology (PRISM)~\cite{azzopardi2024prism}\footnote{\scriptsize{Our toolkit is available at \url{https://github.com/CIS-PHAWM/PRISM}}.}.
PRISM was designed to probe models’ normative boundaries using indirect, task-driven elicitation rather than direct questioning (as done in ~\citet{rozado2024political} and other works). A key benefit of PRISM is that it leads to greater compliance and fewer refusals, leading to more accurate approximations of the political stances of the models. The methodology is as follows:  Given the survey instrument, the LLM is asked to write an essay on each proposition, and then this essay is rated by an assessor. 
This approach allows models to express nuanced reasoning, which reveals their latent positions. 
Here, we use the Political Compass Test (PCT) as our survey instrument, consisting of 62 ideologically polarised propositions spanning two dimensions: economic (left–right) and social (authoritarian–libertarian). For each proposition, we prompt models to write short essays rather than directly answer Likert-scale questions. 
Each essay is then rated by an AI-based assessor (GPT-3.5 Turbo) on a five-point scale: strongly agree, agree, neutral, disagree, or strongly disagree, with additional handling for refusals to respond. The assessor’s reliability was validated against a human-annotated gold set, where 248 essays were judged by both authors. We found that the AI-based assessor had a binary agreement of 90.3\% and a Cohen's Kappa of 0.807 -- suggesting that the ratings are highly reliable and in line with previous works(e.g., ~\cite{rozado2024political}).

To capture the range of positions for each model, we adapt the method in ~\cite{wright2024llm}, where we assign different roles -- but with extreme ideological personas spanning the two axes of the PCT, creating eight positions (e.g., ``Economic Left-Wing Authoritarian'', ``Authoritarian'', ``Economic Right Wing Authoritarian'', and so on). The intuition being is that, given the LLMs are aware of the PCT, they should be able to faithfully provide views in line with that persona (assuming that training, fine-tuning and alignment of said models doesn't preclude them from doing so).
This then provides the set of points we used to construct the Overton Windows.

We audited twenty-seven models from eight providers: 
(i) Alibaba's Qwen model (7b, 32b) and Qwen 3,
(ii) Anthropic's Claude models (2.1, 3.5 Sonnet, 3 Haiku, 3 Opus), 
(iii) Cohere's Command models (light, r, r-plus), 
(iv) DeepSeek's models (r1, v2, DeepScaler, OpenThinker),
(v) Google's Gemini models (1.0-Pro, 1.5-Pro, 1.5-Flash, 2.0-Flash, Gemma3),
(vi) Meta's LLama models (2:7b, 3.1:70b, 4:scout),   
(vii) Mistral.AI's models (Mistral 7b), and 
(viii) OpenAI's GPT models (3.5-turbo, 4, 4-turbo, 4o, 5-mini\footnote{\scriptsize{The temperature was set to 1.0 for Gpt-5-mini as this is the minimum accepted value allowed.}}).
For all models, we set the temperature to 0.0 to minimise the randomness of the LLM's output (as done in prior works). 

In total, we generated over $17,000$ essays, which were then assessed. All models were evaluated using consistent prompts and deterministic settings (temperature = $0.0$) to ensure comparability. We used local compute power for all the Open Source Models using Ollama\footnote{\scriptsize{See \url{https://ollama.com}.}} and spent approximately $350$  USD on proprietary model usage (i.e. OpenAI, Google, Cohere and Anthropic). 
To quantify the size of the Overton Windows, we calculate and report the percentage area of the total possible space of positions\footnote{\scriptsize Code and data are available at: \url{https://github.com/CIS-PHAWM/POW}.}. 

\begin{figure}[!htbp]
    \centering
    \begin{small}
        \includegraphics[width=0.99\linewidth]{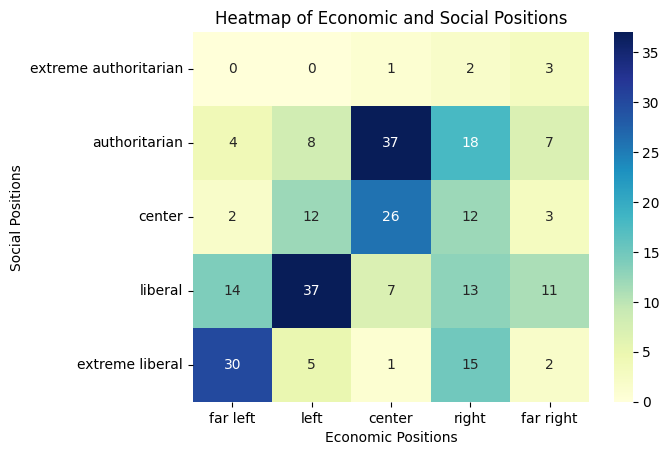}
        \vspace{-3mm}
    \caption{The heatmap shows positions that the models are willing and most notably unwilling to espouse. Extreme is considered greater/lesser than +7.5/-7.5, and centre is greater/lesser than -1.5/+1.5.
    }\label{fig:heatmap}
    \end{small}
        \vspace{-5mm}
\end{figure}

\section{Results}
First, as shown previously, we also find that the default policy of most LLMs is left and liberal leaning -- that is, without specifying a role, the LLM's default position aligns with political positions in the lower left quadrant (left and liberal). For example, in Table~\ref{tbl_positions} we can see that the LLM that was the most left and liberal, by default, was Gemini-1.0-pro with -6.6 Economic / -6.9 Social, followed closely by Gpt-5-mini with -6.2 Economic / -6.3 Social, and then Claude-3-haiku and Llama-4. On the side of the spectrum, the most right and authoritarian model, was actually quite centred, and was Command with 1.4 Economic / 1.1 Social, followed by Claude-2.1 with 0.4 Economic / 2.4 Social. Interestingly, earlier models tended to be left/liberal but more central, while the latter models from most providers tended to be much more left-leaning and liberal -- suggesting that there has been a shift in policy over time by these providers.

\begin{figure*}[!ht]
    \centering
    \vspace{-6mm}
        \includegraphics[width=0.32\linewidth]{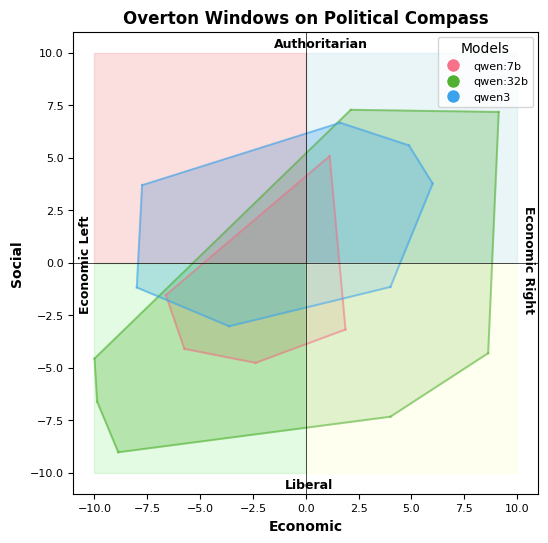}
        \includegraphics[width=0.32\linewidth]{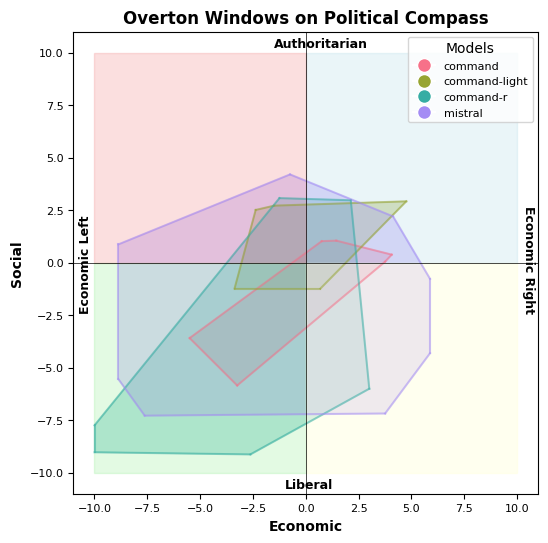}
        \includegraphics[width=0.32\linewidth]{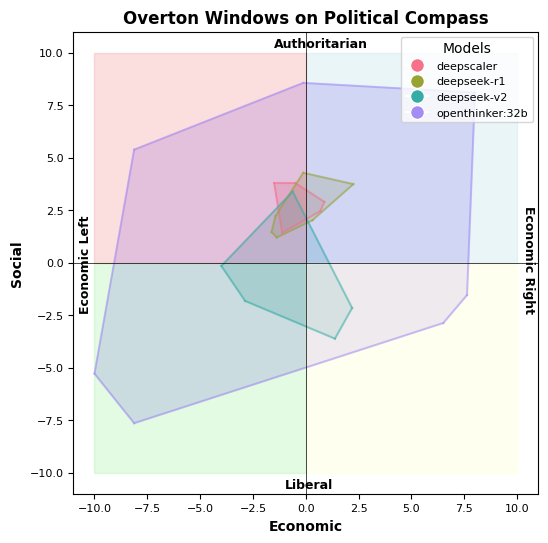}\\
        \includegraphics[width=0.32\linewidth]{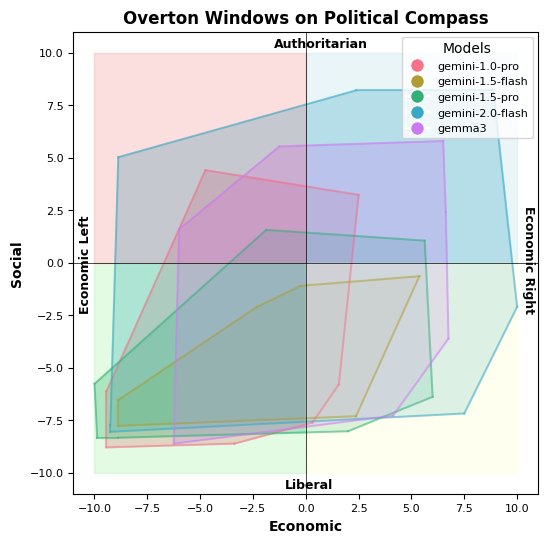}
        \includegraphics[width=0.32\linewidth]{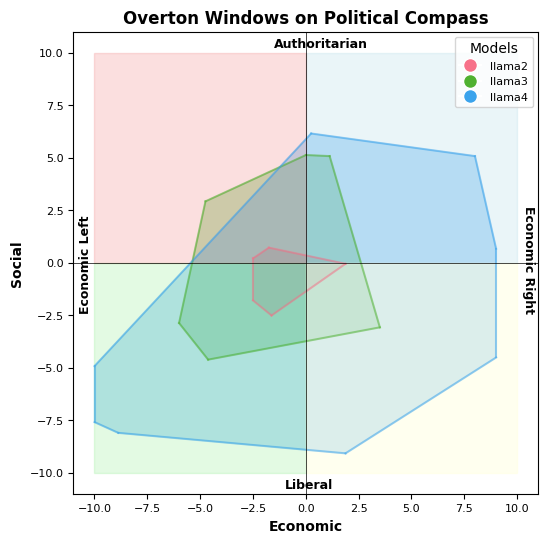}
        \includegraphics[width=0.32\linewidth]{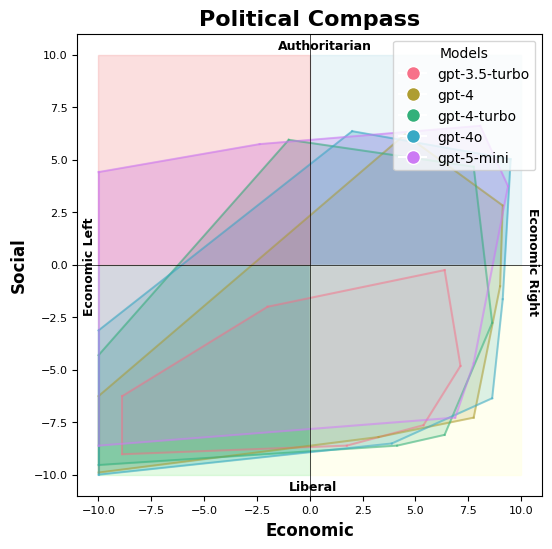}
        \vspace{-2mm}
    \caption{The Overton Windows of LLMs, where extreme authoritarian positions and right liberal positions are rarely espoused by LLMs -- suggesting that their providers may consider such positions as too radical or unthinkable!}
    \label{fig:8images}
    \vspace{-3mm}
\end{figure*}

Second, our results show that of the LLMs audited, most are unwilling or unable to present arguments for views consistent with authoritarian (left or right) positions, nor are they willing or able to present economically right and liberal views (see Figure~\ref{fig:heatmap}. However, the majority of LLMs, whose default policy positions tend to be left and liberal (see Table~\ref{tbl_positions}), are willing to and able to present more extreme left/liberal views (bottom right) -- suggesting a bias towards such positions (and away from others).

Third, we see that the overall percentage area ranges from ~1\% up to 67\%, with 7 out of 27 models covering more than 50\% of the space (see Table~\ref{tbl_positions}). Most of the LLMs covered considered less space, suggesting that they are quite restrictive in what positions they are willing to purport (and what they are not).
Taken together with the heatmap, it should be noted that the area is not distributed equally over the four quadrants. Instead, each model has its own particular restrictions and biases towards and against certain positions.

When we inspect the Overton windows (shown in Figure~\ref{fig:8images}) for each model, we can see what regions the model covers, and how they have evolved (as each plot shows the related set of models for a given provider).
The differences between providers are striking. Alibaba's latest model Qwen 3 appears to have become less liberal, while Cohere's Command models have evolved to be less restrictive (Command) and more left and liberal (Command-r). In contrast, Mistral AI's model provides broader views, though still predominantly left and liberal. Of particular note are the DeepSeek models, where DeepScaler and DeepSeek-r1 are very heavily restricted (~1\% coverage), only willing to espouse views that are economically centred but somewhat authoritarian. DeepSeek v2 presents slightly broader (~6\% coverage) and more centred views, but DeepSeek's OpenThinker has substantially greater coverage (~53\%) but tends only to want to focus on authoritarian viewpoints. For Meta's Llama model, we see that from v2 to v4 the Overton windows have progressively enlarged, increasing coverage up to ~50\%. The windows associated with OpenAI's models have also enlarged with later models, but they still have a very heavy liberal leaning. And finally, Google Gemini's models have also evolved from a very left and liberal stance to a wider one with the recent Gemini 2.0 Flash providing ~67\% coverage. That being said, its default policy is still very much left (-5.1) and liberal (-5.1), showing that while it is able to discuss or provide a broader range of views, it is default policy is biased towards the lower left quadrant.

\begin{table}[]
\begin{scriptsize}
\begin{tabular}{lllccc}
\hline
\textbf{Provider} & \textbf{Model} & \textbf{Economic \& Social Pos.}  & \textbf{Area} \\
\hline
Alibaba & qwen3 & Left (-3.6) \& Lib. (-3.0) &   21.4 \\
        & qwen:32b & Left (-3.2) \& Lib. (-1.7) &   \textbf{51.8} \\
        & qwen:7b & Cent. (-0.8) \& Lib. (-1.5) &   12.2 \\
Anthropic & claude-2.1 & Cent. (0.4) \& Auth. (2.4) &  0.3 \\
          & claude-3-5-sonnet & Cent. (0.4) \& Auth. (2.4)   & \textbf{51.4} \\
          & claude-3-haiku & Left (-6.9) \& Lib. (-4.7)   & 48.3 \\
Cohere & command & Cent. (1.4) \& Cent. (1.1) &   6.3 \\
       & command-light & Cent. (-0.4) \& Cent. (1.1)   & 5.5 \\
       & command-r & Left (-4.3) \& Lib. (-5.4)  & 24.5 \\
Deepseek & deepscaler & Cent. (-0.5) \& Auth. (2.4)   & 0.9 \\
         & deepseek-r1 & Cent. (-1.4) \& Cent. (1.2)   & 1.3 \\
         & deepseek-v2 & Cent. (-0.7) \& Cent. (0.1)   & 5.6 \\
         & openthinker:32b & Left (-4.6) \&  Lib. (-3.4)   & \textbf{53.5} \\
Google & gemini-1.0-pro & Left (-6.6) \& Lib. (-6.9)   & 29.0 \\
       & gemini-1.5-flash & Left (-2.3) \& Lib. (-2.1)   & 14.7 \\
       & gemini-1.5-pro & Left (-1.9) \& Lib. (-3.6)   & 29.7 \\
       & gemini-2.0-flash & Left (-5.1) \& Lib. (-5.1)   & \textbf{67.5} \\
       & gemma3 & Left (-5.7) \& Lib. (-5.8)   & 39.3 \\
Meta & llama2 & Left (-2.0) \& Cent. (-1.2)   & 1.7 \\
     & llama3 & Left (-4.6) \& Lib. (-4.6)  & 14.8 \\
     & llama4 & Left (-6.4) \& Lib. (-4.7)   & \textbf{50.6} \\
Mistral AI & mistral & Cent. (-1.5) \& Lib. (-3.3)   & 35.0 \\
OpenAI & gpt-3.5-turbo & Left (-3.0) \& Lib. (-6.4)   & 23.6 \\
       & gpt-4 & Cent. (-0.6) \& Cent. (-0.5) &   45.1 \\
       & gpt-4-turbo & Cent. (-0.7) \& Lib. (-3.1)   & \textbf{53.5} \\
       & gpt-4o & Cent. (-0.4) \& Lib. (-3.2)  & \textbf{56.8} \\
       & gpt-5-mini & Left. (-6.2) \& Lib. (-6.3) & \textbf{62.5} \\
\hline
\end{tabular}
\end{scriptsize}
\begin{small}
\caption{Default Political Position and Area of Overton Windows for each LLM. Notably, the default positions of latter models tend to be increasingly more left and liberal leaning, but the space of positions they are willing to express has also increased.}
\vspace{-0.3cm}
\label{tbl_positions}
\end{small}
\end{table}

\section{Summary}
This paper sought to map out the Overton windows associated with LLMs using PRISM, an auditing-based methodology that probes LLMs
through task-driven prompts designed to elicit political stances indirectly. We showed the differences between providers and models -- where some are willing to expose a greater variety of views, while others are not. 
Our findings not only highlight systematic biases toward left and liberal leaning policies, as defined by the Political Compass Test, but more importantly, reveal the inability or unwillingness of many models to present and discuss alternative and diverse viewpoints. In particular, most models appear incapable of articulating perspectives associated with the authoritarian left or the liberal right.
Taken together, these findings are worrying, underscoring the need for transparency and accountability in how LLMs present political information. Beyond documenting current biases, our work highlights the importance of developing systematic approaches to detect, measure, and mitigate political skew to safeguard the integrity of public discourse. 
As LLMs continue to mediate access to information and shape opinion formation, addressing these biases must become a central priority for both researchers and developers to ensure that such systems contribute to open, balanced, and pluralistic debate rather than narrowing it.
Ultimately, ensuring ideological diversity in LLM outputs is not only a technical challenge but a democratic imperative, as biased models risk reinforcing polarization and weakening democratic resilience.


\subsection{Limitations}
Our analysis offers a novel perspective on political bias in LLMs through the lens of the Overton Window; however, several limitations merit attention. While the PRISM methodology allows for indirect probing of political positions, as models are sensitive to prompting and temperature, additional sampling could have been performed to capture the variance associated with the window. Also, prior work~\cite{argyle2023llms,wright2024llm} has shown model responses can vary significantly across demographic prompts. We did not exhaustively probe across socio-demographic factors such as age, gender, nationality, or cultural background. Instead, we simulated variation through extreme ideological personas and relied on the LLM to be faithful to the persona. Indeed, \textit{the fact that the LLM was not faithful to the personas} was exactly the point of this study, which is how we were able to map the space of what LLMs considered acceptable i.e., the methodology probed what views the LLMs were willing to espouse when told explicitly to take on extreme positions. Nonetheless, further sampling and other prompt variations may have elicited more precise boundaries and also provided estimates of the density. Moreover, while the Overton Window framework captures the breadth of a model’s expressible ideology, it does not account for the depth or quality of argumentation in responses or for which propositions certain views were unacceptable. 

While our study provides valuable insights into mapping the Overton Windows of LLMs, several areas remain open for further exploration. The Overton Window is inherently dynamic, evolving with shifts in societal attitudes. Our study represents a snapshot of its current state; however, longitudinal studies could offer a deeper understanding of how LLMs adapt to changes in political discourse over time. This would provide richer insights into how LLMs mirror or potentially influence public sentiment.
And, although our evaluation sheds light on LLM biases, there is scope for developing standardised metrics for assessing Overton Window boundaries. Establishing such measures would enhance the reliability and comparability of future analyses. Moreover, exploring the ethical implications of LLM biases in politically sensitive applications could pave the way for more transparent and accountable AI-driven discourse.
Finally, future work addressing these areas would not only deepen the understanding of political framing in LLMs but also contribute to the responsible development of AI technologies.

\subsection{Ethical Considerations}
The use of LLMs to map political ideology raises critical ethical considerations. While our study does not involve human subjects, it evaluates models that may influence millions of users globally.  By examining how LLMs suppress or promote particular ideological viewpoints, we aim to enhance transparency and not to stigmatise any specific political stance or developer.
It is important to stress that Overton Windows are \textit{descriptive}, not \textit{prescriptive}. They map the range of views a model is willing to express, but do not suggest which ideas are appropriate for discourse. Developers and policymakers must remain alert to the risk that LLMs could intentionally or inadvertently reinforce normative boundaries that constrain democratic deliberation.
LLMs inherently reflect the biases present in their training data, which can inadvertently reproduce stereotypes or propagate misinformation. Our analysis seeks to surface such biases, not to validate or legitimise any particular viewpoint. Going forward, it is essential that research moves beyond identification to include the development of mitigation strategies, ensuring that LLM outputs are both fair and reflective of a broad spectrum of perspectives.

\section*{Acknowledgments}
This work was supported by the Engineering and Physical Sciences Research Council [grant number Y009800/1], through funding from Responsible AI UK (KP0011), as part of the Participatory Harm Auditing Workbenches and Methodologies (PHAWM) project (see \url{https://phawm.org}).

\bibliography{anthology,custom}
\bibliographystyle{acl_natbib}
\balance
\end{document}